\documentclass[twocolumn,english,aps,superscriptaddress,prl]{revtex4-1}

\usepackage{array}
\usepackage[latin9]{inputenc}
\usepackage{amsmath}
\usepackage{amssymb}
\usepackage{graphicx}
\usepackage{babel}
\usepackage{mathrsfs}
\usepackage{amsfonts}
\usepackage{epstopdf}
\usepackage{multirow}
\usepackage{color}
\usepackage{natbib}

\begin{document}
\title{Tricritical physics in two-dimensional $p$-wave superfluids}

\author{Fan Yang}
\affiliation{Department of Physics and Astronomy, University of British Columbia, 6224 Agricultural Road, Vancouver, BC, V6T 1Z1, Canada}

\author{Shao-Jian Jiang}
\affiliation{State Key Laboratory of Magnetic Resonance and Atomic and Molecular Physics, Innovation Academy for Precision Measurement Science and Technology, Chinese Academy of Sciences, Wuhan 430071, China}
\altaffiliation{Innovation Academy for Precision Measurement Science and Technology is formerly known as Wuhan Institute of Physics and Mathematics.}

\author{Fei Zhou}
\affiliation{Department of Physics and Astronomy, University of British Columbia, 6224 Agricultural Road, Vancouver, BC, V6T 1Z1, Canada}

\begin{abstract}
We study effects of quantum fluctuations on two-dimensional $p+ip$ superfluids near resonance. In the standard paradigm, phase transitions between superfluids and zero density vacuum are continuous.
When strong quantum fluctuations near resonance are present, the line of continuous phase transitions terminates at two tricritical points near resonance, between which the transitions are expected to be first-order
ones. The size of the window where first-order phase transitions occur is shown to be substantial when the coupling is strong. 
Near first-order transitions, superfluids self-contract due to phase separations between superfluids and vacuum. 
\end{abstract}

\date{\today}
\maketitle

\paragraph{Introduction}
Superfluids and superconductors in $p$-wave channels with potentially topological features have attracted broad interest from condensed matter physics, atomic and molecular physics, and quantum computation communities.
Compared to $s$-wave superfluids, superfluids with $p$-wave pairing embody much richer physics due to more complex manifolds of order parameters as well as possible non-trivial topological invariants and edge modes.
In particular, two-dimensional (2D) superfluids with $p+ip$ pairing support a topological phase \cite{Volovik99,Read00}. Vortex excitations in the topological phase obey non-Abelian statistics \cite{Moore91,Nayak96,Read96,Ivanov01}, which can be further used for quantum computation \cite{Kitaev03,Nayak08}.

$p$-wave superfluidity was observed long ago in liquid ${}^3$He \cite{Leggett75}.
In ultracold Fermi gases, $p$-wave pairing can be realized by preparing fermions in a single pseudo-spin state, so that $s$-wave interactions are {strongly suppressed} by the Pauli exclusion principle. $p$-wave scattering can then be further enhanced by tuning the system to a Feshbach resonance \cite{Chin10} with $p$-wave molecules. 
$p$-wave Feshbach resonance has been observed in ${}^{40}$K and ${}^6$Li gases \cite{Regal03,Zhang04,Schunck05,Gunter05,Gaebler07,Waseem16} despite of particle losses. 
These experiments have further inspired theoretical proposals on potential realizations of $p$-wave superfluids in Fermi gases \cite{Yamaguchi17,Fedorov17}.
In the meantime, progress has been made on theoretical understandings of $p$-wave superfluids at resonance. It is appreciated that various phase transitions, quantum or thermal, between different superfluid phases can further occur near resonance \cite{Botelho05,Gurarie05,Cheng05,Iskin06,Gurarie07}. Most of these studies focus on the limit where quantum fluctuations are perturbative, and the superfluids can be well characterized by mean-field studies.
Furthermore, collisional dynamics and relaxation of molecules in different dimensions have been discussed \cite{Suno03,Levinsen07,Waseem19}.
Fascinating scale symmetric few-body states have also been proposed \cite{Nishida13}.
Very recently, Fermi liquid effects in $p$-wave gases have been further explored \cite{Ding19}.

In the presence of strong  quantum fluctuations, however, superfluids can potentially exhibit highly non-mean-field features due to the breakdown of mean-field theories. 
Therefore, it is necessary to thoroughly examine such possibilities and identify regimes where strong quantum fluctuations in $p$-wave superfluids are dominating.
This is what we intend to achieve in this Letter.
Let us note that there have been a few recent attempts to investigate the role of quantum fluctuations in 2D resonantly interacting $p$-wave Fermi gases \cite{Jiang18,Hu18,Liu18}.
In Ref. \cite{Jiang18}, the authors have illustrated that when quantum fluctuations up to two-loop effects are taken into account, low density homogeneous $p$-wave superfluids can become unstable at resonance. 
In Ref. \cite{Hu18}, fluctuation effects are not included in the effective potential for pairing field dynamics
although fluctuations are taken into account in modifying the number equation. So the renormalization effects studied in Ref. \cite{Jiang18} are beyond the approximation employed in Ref. \cite{Hu18}. 
The variational method in Ref. \cite{Liu18} does show some evidence of instability; however, limitations of the specific numerical scheme remain to be further scrutinized.

\begin{figure}
\includegraphics[width=\columnwidth]{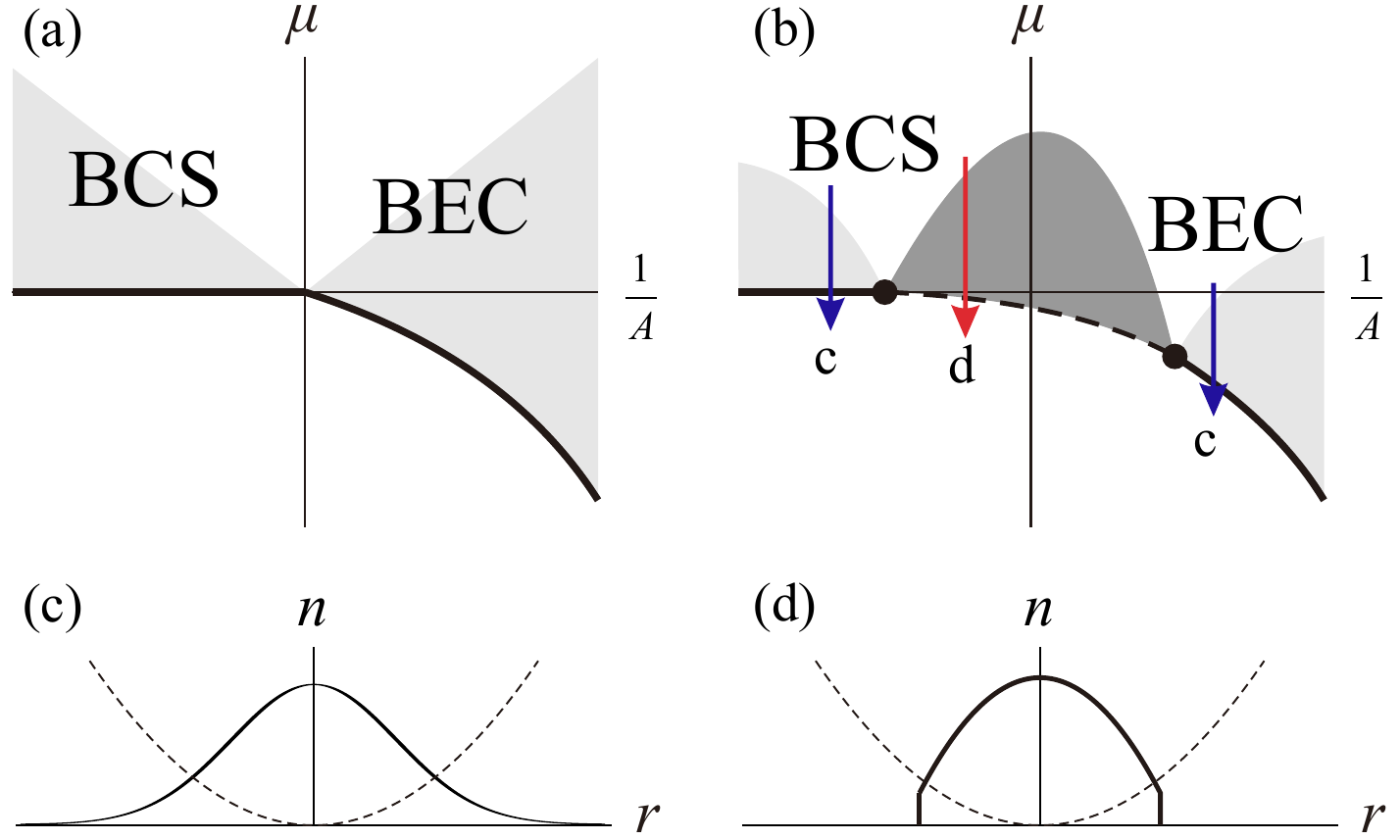}
\caption{
(a) Schematics of superfluid phases in the $\mu$ {\it vs.} $1/A$ plane. 
The horizontal axis is the inverse of $p$-wave scattering area $A$ measuring the detuning. At resonance, $1/A=0$. 
In the standard paradigm, continuous quantum phase transitions occur when $\mu$ reaches a critical value (bold solid line). 
Shaded areas represent weakly interacting regions. 
(b) Near resonance when strong quantum fluctuations are present, the actual transitions become first order (dashed line). Tricritical points (black dots) on each side of resonance terminate the lines of continuous transitions
and separate them from the discontinuous ones. As a result, the weakly interacting regime (light gray) is shifted away from resonance while a strongly fluctuating regime (dark gray) emerges.
In (c) and (d), we illustrate the density profile (solid curve) of superfluids in a trap (dashed curve) away and near resonance respectively.
From the center towards the edge of the trap, the local chemical potential $\mu (r)$ follows the corresponding paths in (b).  \label{demo}}
\end{figure}

\paragraph{Results}
In this Letter, we analyze the consequence of strong quantum fluctuations in 2D $p+ip$ superfluids near resonance. 
We focus on quantum phase transitions between vacuum and the $U(1)$ symmetry-breaking superfluids as chemical potential $\mu$ is tuned across its critical value (Fig. \ref{demo}).
In the standard paradigm, the ground state is a weakly interacting Bardeen-Cooper-Schrieffer (BCS) superfluid on one side of the resonance and a Bose-Einstein condensate (BEC) of diatomic molecules on the other side.
Phase transitions between superfluids and zero density vacuum driven by chemical potentials are therefore expected to be continuous,
and belong to either the free-fermion universality class on the weakly attractive BCS side or the free-boson universality class on the BEC side \cite{transitionline,transitionline1}. 

Effective interactions of the emergent bosonic pairing fields, usually represented by a quartic term in the standard effective potential, always appear to be repulsive at the mean-field level.
Near resonance when quantum fluctuations are strong, we find that these interactions 
are substantially renormalized and can even become attractive.  
The interactions can be shown to change from repulsive to attractive when approaching  resonance from either side.
As a result, continuous phase transitions from vacuum to superfluid phases terminate at a pair of tricritical points located on two sides of resonance as the system is tuned close to resonance. 
Between these tricritical points, quantum transitions driven by chemical potentials are of the first order \cite{Halperin74,Song07,Turner07}, where the particle density jumps from zero to a finite value.
This indicates self-contracted $p+ip$ superfluids near resonance as a consequence of  phase separation between a finite density superfluid and a zero density vacuum.
In trap geometries studied in cold gas experiments, the self-contracted liquid-like superfluids exhibit sharp-edged spatial density profiles, in stark contrast to the conventional smooth Thomas-Fermi density profiles of quantum gases (Fig. \ref{demo}).

We study these phase transitions by employing the following Hamiltonian for 2D $p+ip$ Fermi gases coupled to bosonic molecules,
\begin{align}\label{Hamiltonian}
H=&\sum_{\bf k}\left(\frac{k^2}2-\mu\right)c^\dagger_{\bf k}c_{\bf k}+\sum_{{\bf k}}\left(\frac{k^2}4+\epsilon-2\mu\right)b^\dagger_{{\bf k}}b_{{\bf k}}\nonumber\\
&+\frac {g_0}{\sqrt\Omega}\sum_{{\bf k},{\bf q}}(k_-b_{{\bf q}}c^\dagger_{\frac{\bf q}2+{\bf k}}c^\dagger_{\frac{\bf q}2-{\bf k}}+k_+c_{\frac{\bf q}2-{\bf k}}c_{\frac{\bf q}2+{\bf k}}b^\dagger_{{\bf q}})\nonumber\\
&+\frac{u_3}{\Omega}\sum_{{\bf k,k',q}}c^\dagger_{\frac{\bf q}2-{\bf k}}b^\dagger_{a,\frac{\bf q}2+{\bf k}}b_{\frac{\bf q}2+{\bf k}'}c_{\frac{\bf q}2-{\bf k}'}\nonumber\\
&+\frac{u_4}{\Omega}\sum_{{\bf k,k',q}}b^\dagger_{\frac{\bf q}2+{\bf k}}b^\dagger_{\frac{\bf q}2-{\bf k}}b_{\frac{\bf q}2-{\bf k}'}b_{\frac{\bf q}2+{\bf k}'}.
\end{align}
$c_k^{(\dagger)}$ and $b_{k}^{(\dagger)}$ represent fermions and bosons, respectively. $k_\pm=k_x\pm ik_y$, $\Omega$ is the total area of the system, $\mu$ is the fermion chemical potential,  $\epsilon$ is the bare detuning and $g_0$ is the bare inter-channel coupling.
The bare three- and four-body interactions ($u_3$ and $u_4$) are zero but will be naturally induced in the course of renormalization when quantum fluctuations are integrated out.
To facilitate later discussions on the renormalization group equation (RGE) analysis, we keep these terms explicitly in the Hamiltonian.
For discussions on first order transitions, we will also have to include six-body interaction $u_6$ that is not shown explicitly here.
Note that the detuning from resonance can be more conveniently measured by scattering area $A$ defined as
$A=\left(-\frac\epsilon{g_0^2}+\frac{\Lambda_0^2}{2\pi}\right)^{-1}$.
Resonance occurs when $A\to\infty$, i.e., $\epsilon=g_0^2\Lambda_0^2/2\pi$ with $\Lambda_0$ being the UV cutoff. 

To study phase transitions, we analyze the effective potential $\Phi$ associated with Hamiltonian (\ref{Hamiltonian}) as a function of order parameter $\Delta=g_0\langle b_{0}\rangle/\sqrt\Omega$. The effective potential has the general form
\begin{equation}\label{phi}
\Phi=r|\Delta|^2+V_4|\Delta|^4+V_6|\Delta|^6+...
\end{equation}
$r$, $V_4$ and $V_6(>0)$ are functions of $\mu$, $1/A$ and microscopic parameters like $g_0$, etc. These functions can be conveniently obtained by taking into account of quantum fluctuations via the RGEs.
Following the standard Landau theory \cite{Landau}, the order of phase transitions depends on the sign of $V_4$. When $V_4$ is positive, continuous phase transitions occur at $r=0$; when $V_4$ is negative, first order phase transitions occur at $r=V_4^2/(4V_6)$ with $V_6 >0$. Thus, $r=V_4=0$ corresponds to tricritical points. 
We will employ these elementary relations to identify transitions and tricritical points.

For our model, $r$ is proportional to bosonic chemical potential $\mu_B$ (up to a factor $Z$, see {\it Methods}), with $\mu_B=2\mu$ on the BCS side and $\mu_B=2\mu+W$ on the BEC side, and $W$ being the binding energy. 
If $V_4 $ is positive, continuous phase transitions occur at $\mu_B=0$, leading to the standard paradigm in Fig. \ref{demo}(a).
In the mean-field theory, one can indeed show that $V_4 \propto \ln (\Lambda_0/\Lambda_\text{IR}) > 0 $, implying repulsive interactions of bosonic fields and continuous phase transitions. 
Here $\Lambda_\text{IR} (\ll \Lambda_0)$ is an infrared (IR) scale relevant to superfluids and will be specified later.
Far away from resonance, quantum fluctuations can be treated perturbatively, and $V_4$ remains positive. 
However, we find that quantum fluctuations become very strong near resonance and substantially renormalize $V_4$. 
As a result, $V_4$ can become negative and phase transitions become of the first order.

\begin{figure}
\includegraphics[width=0.65\columnwidth]{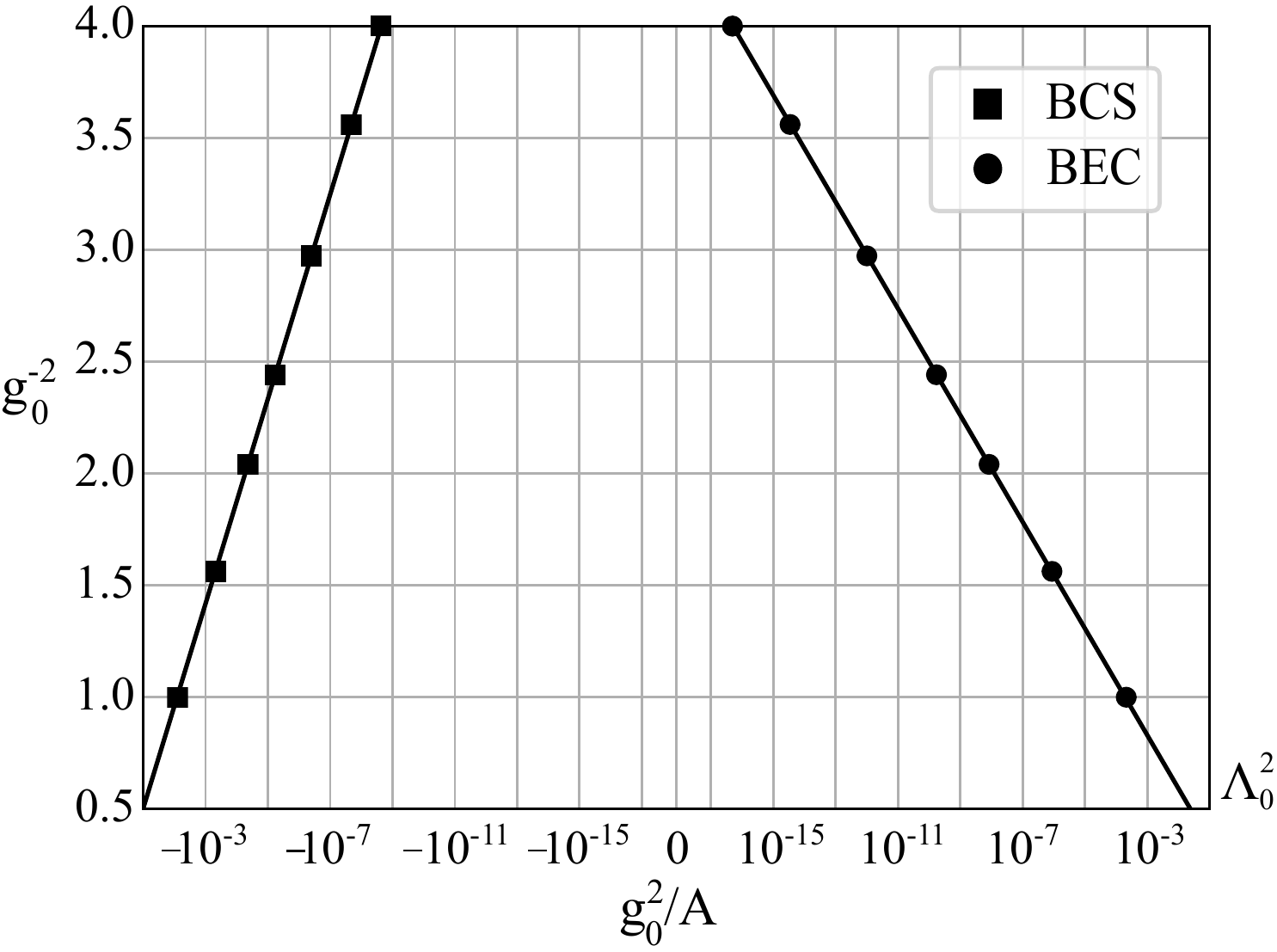}
\caption{Tricritical points for different $g_0$ are labeled by squares and dots on the BCS and BEC side, respectively. They are fitted to Eq. (\ref{A}) represented by the straight lines. Phase transitions 
driven by chemical potentials are first order between the tricritical points, and continuous otherwise. \label{detuning_fit}}
\end{figure}

To identify tricritical points separating the continuous and first order transitions,
we compute, for given $A$ and $g_0$, $V_4 (A)$ along the $\mu_B=0$ transition line by numerically solving a set of coupled RGEs (see Methods). 
From the data of $V_4$, we pinpoint the value of $A^*$ at which $V_4 (A)$ changes from repulsive ($V_4 >0$) to attractive ($V_4 <0$).
The values of $A^*$ are then identified as tricritical points that terminate the continuous transitions when approaching resonance; beyond these points, the system undergoes first order transitions. 
We plot these tricritical points $A^*$ for different $g_0$ and their scaling behavior in Fig. \ref{detuning_fit}.
Along the transition line $\mu_B=0$ but away from resonance ($|A|<|A^*|$), the phase transition is continuous with $V_4 (A)>0$;
while close to resonance ($|A|>|A^*|$), $V_4 (A)$ is negative and the phase transition is first order.
For small and even intermediate two-body couplings $g_0\lesssim1$, the general structure of the RGEs  suggests the following scaling behavior of $A^*(g_0)$
\begin{equation}\label{A}
|A^*(g_0)|^{-1}\sim g_0^{-2} e^{-c/g_0^2}.
\end{equation} 
Numerically, we find that $c\approx5.035$ on the BCS side and $c\approx9.662$ on the BEC side.
Eq. (\ref{A}) is one of the central results of our analyses.
When $g_0 \to 0$, the tricritical points effectively merge at resonance as $A^*\to\infty$ on both sides, and the window of first order transitions vanishes, a characteriztic of the mean-field theory. 
For small but finite values of $g_0$, the window of the first order phase transition is exponentially small but it becomes very substantial when $g_0$ increases.

\begin{figure}
\includegraphics[width=\columnwidth]{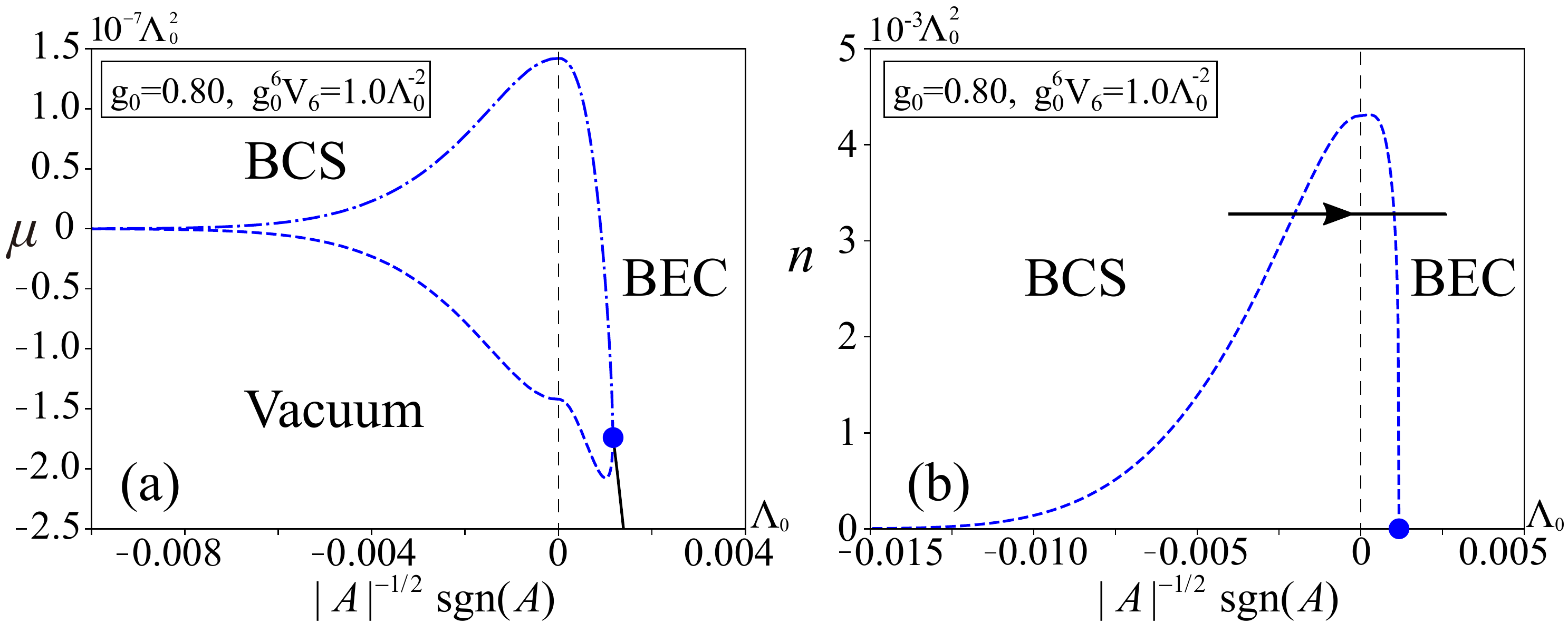}
\caption{(a) Phase diagram in the $\mu$ {\it vs.} $1/A$ plane. Vertical dashed line marks resonance.  First order phase transition line $\mu_c (A)$ (lower dashed curve) is separated from continuous phase transition line (solid curve) by a tricritical point (dot) on each side of resonance.  (The Tricritical point on BCS side is outside the range of this plot.)  
The upper dotted-dashed curve represents the crossover $\mu^*(A)$ between strong and weak fluctuation regimes. Below the crossover line, quantum fluctuations dominate in the superfluid phase.  (b) Phase diagram mapped to $n$ {\it vs.} $1/A$ plane from (a) using equation of state (\ref{eos}).  The dashed curve represents the critical density $n_c(A)$ in first order transitions, below which superfluids are self-contracted liquid-like droplets with constant density $n_c$. Note that at a fixed low density as the scattering area is varied continuously, superfluids can undergo two consecutive transitions as illustrated by the horizontal trajectory.\label{PD}}
\end{figure}

To further determine the first order transition line $\mu_c (A)$ between tricritical points, we numerically solve, for a given $A$, the equation of transition condition $r(\mu_c) =V^2_4(\mu_c)/(4V_6)$ using the data of $V_4$ and $V_6$ obtained from the RGEs. The resultant first order transition line is presented in Fig. \ref{PD}(a). 
Next, we extend our analysis to superfluids above this transition line. 
In Fig. \ref{PD}(a), we also present the fluctuation dominant region in the phase diagram by computing the chemical potential $\mu^*(A)$, at which $V_4(\mu^*)=0$ for a given scattering area $A$.
$V_4$ is positive for $\mu>\mu^*(A)$, where mean-field results remain qualitatively correct and quantum fluctuations do not play a very significant role.
$V_4$ is negative for $\mu<\mu^*(A)$, where quantum fluctuations strongly renormalize the interactions and the physics is dominated by strong fluctuations.
The line of $\mu^*(A)$ represents a crossover between the strongly and weakly fluctuating regimes.

To understand superfluids with a given density, we map the phase diagram onto the density plane in Fig. \ref{PD}(b) using the equation of state.
For continuous phase transitions, homogeneous superfluids exist at an arbitrarily low density. 
When phase transitions are of the first order, fermion density changes discontinuously at critical chemical potential $\mu_c (A)$.
Consequently, homogeneous superfluids can only exist with densities higher than a threshold $n_c(A)$. In other words, at $n<n_c (A)$ superfluids are self-contracted quantum liquids.
Practically, at a fixed low density when approaching resonance, a superfluid first undergoes a phase separation from vacuum forming a self-contracted droplet at a critical scattering area before reaching resonance;
after crossing resonance, it makes another transition back to a homogeneous state at another critical scattering area as shown in Fig. \ref{PD}(b). 
The self-contracted superfluids or phase separation process can be considered as a clear manifestation of attractive interactions between bosonic pairing fields due to strong fluctuations.

We compute the equation of state from the effective potential via the relation $n=-\partial\Phi/\partial\mu$ and $\partial\Phi/\partial|\Delta|=0$. 
We obtain the critical density $n_c (A)$ at the first order transition line (i.e., at $\Phi(\Delta)=0$): 
\begin{equation}\label{eos}
n_c \approx2\sqrt{\frac{|\mu^c_B|}{V_6g_0^6}}\left(1+\frac{g_0^2}\pi\ln\frac{\Lambda_0}{\sqrt{|\mu^c_B|}}\right)^{3/2},
\end{equation}
where $\mu^c_B$ is the chemical potential for the bosonic pairing field along the first order transition line $\mu_c (A)$ \cite{crossover}.

\paragraph{Methods}
In the effective potential (\ref{phi}), $r$, $V_4$ and $V_6$, which depend on various quantum fluctuations, are directly related to $\mu_B$, and interactions $u_4$ and $u_6$. 
To establish the relations, we consider a Hamiltonian defined at scale $\Lambda=\Lambda_0 e^{-s}$, with scale dependent parameters,  $\epsilon(s)$, $u_4(s)$, $u_6(s)$, etc, so that Hamiltonians defined at different scales lead to the same effective potential.
These scale dependent parameters then follow the standard RGEs that effectively take into account renormalization effects due to quantum fluctuations \cite{Coleman73,Peshkin}.  

In the IR limit, the solution to the RGEs of these parameters can directly yield the values of  $r$, $V_4$ and $V_6$.
For instance, at  scale $\Lambda_\text{IR}=\Lambda_0e^{-s_{\text{IR}}}$ relevant to our effective theory (see details below),
$r=-\mu_B g_0^{-2}[Z(\Lambda_\text{IR})]^{-1}$,
$V_4=u_4(\Lambda_\text{IR})g_0^{-4}[Z(\Lambda_\text{IR})]^{-2}$,
and $V_6=u_6(\Lambda_\text{IR})g_0^{-6}[Z(\Lambda_\text{IR})]^{-3}$.
The field renormalization $Z(\Lambda_\text{IR})$ plays an important role in our discussions because  the effective potential is expressed in terms of $\Delta$ proportional to bare fields defined at UV scale $\Lambda_0$.

On the BEC side ($A>0$), the RGEs along the continuous transition line ($\mu_B=2\mu+W=0$) are
\begin{align}\label{u4BEC}
\frac{du_4}{ds}=-\frac{2g^2u_4}{\pi(1+\tilde W)}+\frac{4g^4}{\pi(1+\tilde W)^3}
+\frac{2g^2u_3}{\pi(1+\tilde W)^2}-\frac{2u_4^2}\pi,
\end{align}
\begin{equation}\label{Z}
\frac{d\tilde u_6}{ds}=-2\tilde u_6-\frac{3g^2\tilde u_6}{\pi(1-2\tilde\mu)}, \quad \frac{d\ln Z}{ds}=-\frac{g^2}\pi\frac1{1-2\tilde\mu},
\end{equation}
\begin{equation}
\frac{d\tilde W}{ds}=2\tilde W, \qquad \frac{dg}{ds}=-\frac{g^3}{2\pi}\frac1{1-2\tilde\mu},\qquad \frac{d\tilde\mu}{ds}=2\tilde\mu,  
\end{equation}
\begin{equation}
\frac{du_3}{ds}=-\frac{16g^4}{3\pi(1+\tilde W)^2}-\frac{11g^2u_3}{3\pi(1+\tilde W)}-\frac{2u_3^2}{3\pi}.
\end{equation}
Here, we define dimensionless quantities $\tilde\mu=\mu/\Lambda^2$, $\tilde W=W/\Lambda^2$ and $\tilde u_6=u_6\Lambda^2$. Note that binding energy $W$,  instead of microscopic detuning $\epsilon(s)$, appears in the RGEs when the system is off resonance.
And we have kept the most relevant field renormalization effect in $u_6$.

On the BCS side ($A<0$), although there are no physical bound states, quasi-binding energy $W'<0$ appears in the RGEs.
The RGEs of $u_4$ and $u_3$ on the BCS side at the continuous transition line  ($\mu_B=2\mu=0$) take the following form, 
\begin{equation}\label{u4BCS}
\frac{du_4}{ds}=-\frac{2g^2u_4}\pi+\frac{4g^4}\pi+\frac{2g^2u_3}\pi-\frac{2u_4^2}{\pi(1-4\tilde W')},
\end{equation}
\begin{equation}
\frac{du_3}{ds}=-\frac1{3-4\tilde W'}\left(\frac{16g^4}{\pi}+\frac{8g^2u_3}{\pi}+\frac{2u_3^2}{\pi}\right)-\frac{g^2u_3}\pi.
\end{equation}
Here $\tilde W'=W'/\Lambda^2$ satisfies $d\tilde W'/ds=2\tilde W'$.
Binding energy $W$ and quasibiding energy $W'$ are both related to the scattering area by the same equation
\begin{equation}\label{DA}
W^{(\prime)}\left(1+\frac{g_0^2}\pi\ln\frac{\Lambda_0}{\sqrt{|W^{(\prime)}}|}\right)=\frac{g_0^2}A.
\end{equation}

In the above RGEs, the mean-field effect is represented by the first two terms in Eqs. (\ref{u4BEC}) and (\ref{u4BCS}). 
One can verify that at resonance these two terms lead to $u_4^{\text{MF}}(s)=4g_0^4s/[\pi(1+g_0^2s/\pi)^{2}]$,
which yields a positive definite $V_4^\text{MF}=(4/\pi)\ln(\Lambda_0/\Lambda_\text{IR})>0$, a standard mean field result.

At resonance ($W=W'=0$) and zero chemical potential, the RG equations are identical to those derived for few-body studies in Ref. \cite{Nishida13}.
Compared to Ref. \cite{Nishida13}, we have further included two new ingredients in the RGEs to facilitate our many-body studies:
first, we allow a finite detuning to address off-resonance physics; second, we introduce finite chemical potentials to address the properties of superfluids.

We first numerically solve the RGEs at the transition line $\mu_B=0$, where the IR scale is $\Lambda_\text{IR}=0$ (i.e. $s_\text{IR}\to\infty$) \cite{r}. 
$u_4(s)$ in this limit can be shown to have a very simple scaling form, $su_4(s)=f(g^2_0 s; g^2_0 s_W, g_0^2)$ where $s_W=(1/2)\ln (\Lambda^2_0/|W^{(\prime)}|)$ and $f$ is a universal function.
We compute $su_4(s)$ for large $s$ and identify the tricritical points by computing the critical scattering area $A^*$ where $s_\text{IR}u_4(s_\text{IR})=0$ as $s_\text{IR}\to\infty$.
The above scaling form indicates that at tricritical points, $f(\infty; g^2_0 s_W, g_0^2)=0$, which yields $g_0^2 s_W=\frac c2+ c_1g_0^2+...$ when $g_0$ is small \cite{AF}. This leads to the scaling relation in Eq. (\ref{A}). 

To obtain the first order transition line, we utilize the relations between $\mu_B$, $u_4(s_\text{IR})$, $u_6(s_\text{IR})$ 
and $r$, $V_4$, $V_6$, 
and numerically obtain $\mu_c (A)$ from the transition condition $r(\mu_c) =V^2_4(\mu_c)/(4V_6)$ at a given scattering area $A$.  Here the IR scale is set to be $s_\text{IR}^c=(1/2)\ln(\Lambda_0^2/|\mu_B^c|)$.
For the crossover line,  we solve the same set of RGEs. From the solutions to RGEs, we compute $\mu^*(A)$ that satisfies the crossover condition $V_4(\mu^*)=0$, which requires $s^*_\text{IR}u_4(s^*_\text{IR})=0$ at the IR scale $s_\text{IR}^*=(1/2)\ln(\Lambda_0^2/\mu_B^*)$.

\paragraph{Discussions} 
Although the RGEs in this Letter work best for $g_0\lesssim1$, it can still offer a valuable insight of qualitative properties of superfluids with strong interactions ($g_0\gg1$). 
If we extrapolate our results to large $g_0$, the window of first order transitions becomes very significant, even though we do not expect our theory to be quantitatively accurate in this limit.

Quantum fluctuations can, in principle, further stabilize other competing states or orders.
One such possibility is  to form polar states, i.e. $p_x$- or $p_y$-type of superfluids, which break the $U(1$) and rotational symmetries, but unlike $p+ip$ superfluids, do not break the time-reversal symmetry. 
We have applied the RGEs to further examine interactions between $p+ip$ and $p-ip$ fields. Close to tricritical points, we find no evidence of $p_x$- or $p_y$-type of ordering. 
In this Letter, we have exclusively focused on superfluids and pairing states, which shall be most relevant if one simply follows the lower branch starting from the BCS side. At resonance ($A\to\infty$), there can also exist other more subtle few-body clusters as proposed in Ref. \cite{Nishida13}. Applying the RGEs to tricritical points and transition lines $\mu_c(A)$, we have found no numerical evidence of few-body cluster states near $A^*$. 
However, whether or not tricritical points discussed here can be related to a precursor of those more subtle non-pairing states is still an open question; 
interplays between superfluids and other exotic states and detailed relaxation dynamics beyond pairing physics remain to be investigated in the future.

\begin{acknowledgments}
\paragraph{Acknowledgments}
We thank V. Gurarie, H. Hu and S. Moroz for helpful discussions.
This project is in part supported by Canadian Institute for Advanced Research.  
S.-J. Jiang would like to acknowledge the support by National Natural Science Foundation of China (Grant No. 11804376).

\end{acknowledgments}


\begin{thebibliography}{References}
\bibitem{transitionline} The transitions are characterized by free-fermion or free-boson field theories while interactions are renormalized away in the infrared limit.
\bibitem{crossover} Here we have take into account of field renormalization $Z$. The crossover line can also be mapped by computing the equation of states at $\mu^* (A)$ instead of $\mu_c (A)$. 
Density $n^* (A)$ along the crossover line turns out to be close to the critical density $n_{c} (A)$ at the first order transition line and 
hence $n_c$ also yields an estimate of $n^*$.
\bibitem{AF}$f(\infty; x=g^2_0 s_W, y=g_0^2)$ is a smooth function in the limit of $y=0$.
\end{thebibliography}
\end{document}